\documentstyle[aps,twocolumn]{revtex}

\begin{document}
\title{Spontaneous Fluxon Formation in \\
Annular Josephson Tunnel Junctions\thanks{%
Submitted to Phys. Rev. B.}}
\author{R. Monaco$^{a}$\thanks{%
E-mail: roberto@sa.infn.it}, J.\ Mygind$^{b}$\thanks{{E-mail: }%
myg@fysik.dtu.dk} and R.\ J.\ Rivers$^{c}$\thanks{%
E-mail: r.rivers@ic.ac.uk, Permanent address:\newline
Imperial College, London, SW7 2BZ, U.K.}}
\address{{\it a}) Istituto di Cibernetica del C.N.R., I-80078, Pozzuoli, Italy\\
and Unita' INFM-Dipartimento di Fisica, Universita' di Salerno,\\
I-84081 Baronissi (SA), Italy\\
b) Department of Physics, Technical University of Denmark, \\
B309, DK-2800 Lyngby, Denmark\\
{\it c}) Centre of Theoretical Physics, University of Sussex, \\
Brighton, BN1 9QJ, U.K. }
\date{\today }
\maketitle

\begin{abstract}
It has been argued by Zurek and Kibble that the likelihood of producing
defects in a continuous phase transition depends in a characteristic way on
the quench rate. In this paper we discuss our experiment for measuring the
Zurek-Kibble scaling exponent $\sigma $ for the production of fluxons in
annular symmetric Josephson Tunnel Junctions. The predicted exponent is $%
\sigma =0.25$, and we find $\sigma =0.27\pm 0.05$. Further, there is
agreement with the ZK prediction for the overall normalisation.

\end{abstract}

\section{Introduction}

As the early universe cooled it is believed to have undergone a series of
spontaneous phase transitions, whose inhomogeneities (monopoles, cosmic
strings, domain walls) have observable consequences, for structure formation
in particular. These defects appear because the correlation length $\xi $ of
the field (or fields) whose expectation value is the order parameter is
necessarily {\it finite} for a transition that is implemented in a finite
time, whether it be continuous or not.

It is difficult to determine the microscopic dynamics of such fields but,
using only simple causal arguments Kibble\cite{kibble1,kibble2} made
estimates of this early field ordering, and the density of topological
defects produced at Grand Unified Theory (GUT) transitions at $10^{-35}s$.
Unfortunately, because the nature of the field theories is not known with
any reliability, and the effects of their evolution are not visible until
the decoupling of the radiation and matter $10^{6}yrs$ later, it is
impossible to provide unambiguous checks of these predictions. However,
causality is such a fundamental notion that Zurek suggested\cite
{zurek1,zurek2} that identical causal arguments, with similar predictions,
were applicable to condensed matter systems for which direct experiments on
defects could be performed. In addition to their intrinsic interest for a
better understanding of the dynamics of transitions in condensed matter, the
hope is that successful tests of these predictions can lead to a better
understanding of phase transitions in quantum fields.

Several experiments in condensed matter systems have already been performed%
\cite{helsinki,grenoble,lancaster,lancaster2,technion,pamplona,florence} to
test the Zurek-Kibble predictions, with mixed results. It is these
predictions that we have tested here, using annular Josephson tunnel
junctions, for which the defects are {\it fluxons}.

This paper is organised as follows. In the next section we discuss the
Zurek-Kibble (ZK) scenario. Successive sections then give the predictions
for annular Josephson Tunnel Junctions (AJTJs), show how the fluxons are
measured, the nature of the samples, describe the experimental setup and,
finally, present the measurements and their agreement with the predictions.
An abbreviated description of the experiment has been given elsewhere\cite
{MMR}. Details of the theoretical analysis can be found in \cite{KMR,MRK},
in which an earlier experiment\cite{Roberto} by two of us (R.M and J.M) was
analysed, to demonstrate its compatibility with the ZK analysis, even though
it had not been performed with this in mind.

\section{Zurek-Kibble Causality}

Consider a system with critical temperature $T_{c}$, cooled through that
temperature so that, if $T(t)$ is the temperature at time $t$, then $%
T(0)=T_{c}$. ${\dot{T}}(0)=-T_{c}/\tau _{Q}$ defines the quench time $\tau
_{Q}$.

There are several ways\cite{zurek2} of formulating the Zurek-Kibble
causality bounds, but they all depend on the fact that, as the transition
begins to be implemented, there is a maximum speed $c(t) = c(T(t))$ at which
the system can become ordered. For relativistic quantum field theory (QFT) $%
c $ is the speed of light. For superfluids, $c(t)$ is the speed of second
sound, vanishing at $t=0$. For Josephson tunnel Junctions $c(t)$, which
depends on the nature of the junction, is the Swihart\cite{Swihart} velocity.

Suppose that the equilibrium (adiabatic) correlation length $\xi
_{ad}(t)=\xi _{ad}(T(t)) $ diverges near $t=0$ as 
\[
\xi _{ad}(t)=\xi _{0}\bigg|\frac{t}{\tau _{Q}}\bigg|^{-\nu }. 
\]
\noindent However, the true non-equilibrium correlation length $\xi (t)$ can
only change so much in a finite time, and does not diverge. Kibble and Zurek
made two assumptions:

\begin{itemize}
\item  Firstly, the correlation length ${\bar{\xi}}$ of the fields that
characterizes the onset of order is the equilibrium correlation length ${%
\bar{\xi}}=\xi _{ad}({\bar{t}})$ at some appropriate time ${\bar{t}}$.

\item  Secondly, we can measure ${\bar{\xi}}$ experimentally by measuring
the number of defects, assuming that the defect separation $\xi _{def}=O({%
\bar{\xi}})$.
\end{itemize}

There are several ways to estimate ${\bar t}$ (the 'causal time'), explicit
or implicit in the early work of Zurek\cite{zurek2}. Most simply,

\begin{itemize}
\item  $\xi (t)$ cannot grow faster than $c(t)$. This is true both before
and after the transition. That is, $\bar{t}$ is defined by the condition
that ${\dot{\xi}}_{ad}({\bar{t}})\approx -c({\bar{t}})$.

\item  The relaxation time for the long wavelength modes $\tau (t)$ is
defined by $c(t)=\xi _{ad}(t)/\tau (t)$. From this viewpoint ${\bar{t}}$ is
that time when we can return to an adiabatic regime, ${\bar{t}}\approx \tau (%
{\bar{t}})$.
\end{itemize}

In simple systems these estimates agree, up to numerical factors
approximately unity\cite{zurek2}. They give ${\bar{t}}$ of the form ${\bar{t}%
}\sim\tau _{Q}^{1-\gamma }\tau _{0}^{\gamma }$, where $\tau _{0}\ll \tau
_{Q} $ is the cold relaxation time of the longest wavelength modes, and the
critical exponent $\gamma $ depends upon the system. As a result, $\tau
_{Q}\gg {\bar{t}}\gg \tau _{0}$. Identifying the initial domain size and
defect separation as in the second ZK assumption then gives 
\begin{equation}
{\bar{\xi}}\sim \xi _{ad}({\bar{t}})=\xi _{0}\bigg(\frac{\tau _{Q}}{\tau _{0}%
}\bigg)^{\sigma }\gg \xi _{0},  \label{xibar}
\end{equation}
where $\sigma =\gamma \nu $. This is very large on the scale of cold defects
which shrink to size $\xi _{ad}(T_{fin})= O(\xi_0)$, where $T_{fin}$ is the
final temperature, and $\xi _{0}$ is determined from the microscopic
dynamics. We term $\sigma $ the Zurek-Kibble (ZK) scaling index.

The analysis above is for large systems, of linear size $L\gg{\bar\xi}$. For
the experiment that we shall describe below, of Annular Josephson Tunnel
Junctions (AJTJs) of circumference $C$, we find that is $C<{\bar\xi}$. In
fact, we expect the conclusions to be equally valid for small systems, for
which the relevant time might, incorrectly, seem to be the time when the
coherence length becomes smaller than the system. The reason is that the
causal bounds (\ref{xibar}) are to be thought of as a short-hand for the
underlying dynamics. At the microscopic level, causality along the lines
above is not explicit, although encoded in the relevant dynamical equations.
The picture is rather one of order being established through the growth of
the amplitudes of long-wavelength instabilities. The earliest time at which
we can identify defects from this viewpoint is when the order parameters
have achieved their equilibrium magnitudes. Qualitatively, for simple models
this time is in good agreement with the causal time ${\bar t}$ above. There
is no real surprise in this. It has been shown by one of us\cite{RKK,RR}
that, in general, the causal time and distance scales ${\bar t}$ and ${\bar%
\xi} = \xi({\bar t})$ are just as we would expect from dimensional analysis
(in the mean-field approximation) and unstable modes grow exponentially,
whereby the dependence of the causal time (and corresponding defect density)
on the microscopic parameters is only logarithmic. In the same way, the
distance between field zeroes has the same scaling dependence on $\tau_Q$ as 
${\bar\xi}$ of (\ref{xibar}), up to logarithms.

As for the production of defects, they are transitional regions between
different system ground-states. For superfluids like $^4He$ and
superconductors these transitional regions are flagged by {\it zeroes} of
the scalar order-parameter field. For the case of JTJs we shall see that
this is generalised to {\it zeroes (mod $2\pi$)} of the order-parameter
field. Any field crossing zero (mod $2\pi$) has the potential to mature into
a defect. However, only when the transition is complete will the field
configurations in the vicinity of the zeroes have the energy profile of a
classical defect, the solution to the classical field equations. Thus,
before the causal time we now have a picture in which there is a fractal
thermal fuzz of potential defects, whose density depends on the scale at
which we look. By the causal time some of these have developed into the
(scale-independent) defects that we see subsequently (see \cite{RKK,RR}).
Because, from this viewpoint, we have many proto-defects jockeying to become
the real thing, the relevant scale to compare to the system size is not ${%
\bar\xi}$ but $\xi_0$, as before. Equivalently, counting field zeroes (mod $%
2\pi$, or not) depends on the short-distance behaviour of the field
correlations. We do not expect a problem as long as $C\gg\xi_0$, as is
always the case.

We conclude by noting that this is a very different picture from that of
domains freezing in as the transition is approached from {\it above} ($t<0$%
), which is how causal bounds were originally posed\cite{kibble2,zurek1}.
What matters is that all these causal descriptions give results with the
correct engineering dimensions. With this in mind, we keep the causal bound (%
\ref{xibar}) as a convenient mnemonic.

\section{The ZK Predictions For Fluxons}

The order parameter of a Josephson tunnel junction at temperature $T<T_{c}$
is the phase difference $\phi $ of the macroscopic superconducting quantum
mechanical wave functions across the barrier. Using a Lagrangian formalism
Gr\o nbech-Jensen {\it et al.} \cite{lom} have shown that, for an annular
JTJ with a distributed bias current $\gamma $, $\phi $ obeys to the
following perturbed Sine-Gordon equation (PSGE): 
\begin{eqnarray}
\frac{\partial ^{2}\phi }{\partial x^{2}} &-&\frac{1}{c^{2}(T)}\frac{%
\partial ^{2}\phi }{\partial t^{2}}-\frac{1}{\lambda _{J}^{2}(T)}\sin \phi 
\nonumber \\
&=&\gamma +\frac{\alpha }{c^{2}(T)}\frac{\partial \phi }{\partial t}-\beta 
\frac{\partial ^{3}\phi }{\partial x^{2}\partial t}  \label{SG}
\end{eqnarray}
provided the width $\Delta r$ of the annulus, of radius $r$, satisfies $%
\Delta r\ll r$ and $\Delta r\ll \lambda _{J}(T)$, the Josephson coherence
length. In this case $x$ measures the distance along the annulus, and $c(T)$
is the Swihart velocity. $\alpha $ and $\beta $ are the coefficients of the
losses due to the tunnelling current and to the surface impedance,
respectively.

The boundary conditions for Eq.(\ref{SG}) are periodic \cite{scott} and
derive from fluxoid quantization \cite{noi}: $\phi (x+C)=\phi (x)+2\pi n,$
where $C=2\pi r$ is the circumference of the junction and the winding number 
$n$ is an integer corresponding to the algebraic sum of fluxons trapped in
the junction barrier at the normal-superconducting (N-S) transition; $n$ is
a topological system constant, that is, only fluxon-antifluxon ($F\overline{F%
}$) pairs can be created or annihilated as long as the junction remains in
the superconducting state.

The classical fluxons are the 'kinks' of the Sine-Gordon theory. As with
other models of defect formation, equation (\ref{SG}) is only valid once the
transition is complete: therefore, we shall not use it to study the
appearance of fluxons. However, it is sufficient to enable us, in the spirit
of the Zurek-Kibble scenario, to identify $\lambda _{J}(T)$, diverging at $%
T_{c}$, as the equilibrium correlation length $\xi _{ad}(T)$ to be
constrained by causality. Further, the Swihart velocity $c(T)$ (with
critical slowing down at $T=T_{c}$), measures the maximum speed at which the
order parameter can change\cite{Swihart,Barone}.

A detailed discussion of the ZK bounds has been given elsewhere by us\cite
{KMR,MRK}, and we refer the reader to these articles for more details. The
JTJs in our experiment are {\it symmetric}, by which is meant that the
electrodes are made of identical superconducting material with the same
energy gaps and the same $T_{c}$. For such JTJs \cite{KMR,MRK} $\gamma =\nu
=1/2$. Therefore, at the time of their formation 
\[
{\bar{t}}=\sqrt{\tau _{Q}\tau _{0}}, 
\]
the separation of fluxons is expected, in the ZK picture, to be 
\begin{equation}
{\bar{\xi}}\sim \xi _{0}\bigg(\frac{\tau _{Q}}{\tau _{0}}\bigg)^{1/4},
\label{JTJ}
\end{equation}
where $\xi _{0}$ and $\tau _{0}$ have the same meaning as before. It is the
prediction (\ref{JTJ}) that will be tested in our experiment.

In terms of the parameters of the JTJs, $\xi _{0}$ has been inferred\cite
{MRK} as 
\[
\xi _{0}=\sqrt{\frac{\hbar }{2e\mu _{0}d_{s}aJ_{c}(0)}}. 
\]
where $J_{c}(T)$ is the Josephson current density at temperature $T$. The
parameter $a$ is given in terms of the superconductor gap energy and
critical temperature and has a value between $3$ and $5$. If the thickness
of the two superconducting electrodes differs, the effective thickness $%
d_{s} $ is the harmonic mean of the individual thicknesses\cite{Barone}. As
for $\tau _{0}$, it is given as $\tau _{0}=\xi _{0}/c_{0}$, where $c_{0}$
defines the behavior $c(t)=c_{0}(t/\tau _{Q})^{1/2}$of the Swihart velocity
for the system near $T=T_{c}$.

\section{Measuring Fluxons}

Once fluxons have appeared Eq.\ref{SG} is relevant. A consequence of the
periodic boundary conditions in AJTJs is that fluxons behave as relativistic
particles on an infinite lossy line. In the absence of any current through
the barrier and/or externally applied magnetic field the fluxons experience
a flat potential and therefore are in indifferent equilibrium as far as the
barrier is homogeneous and pin-hole free; in the reality, the barrier
defects act as small pinning or repulsive potentials for the fluxons. Due to
the losses, after the transient regime is over the fluxons are still.
Unfortunately static fluxons are difficult to reveal, since according to the
second Josephson law, any static phase profile $\phi _{t}(x,t)=0$ does not
alter the junction zero-voltage state that is also typical of a flat profile 
$\phi (x,t)=const.$, corresponding to the absence of any trapped fluxon.

In contrast, whenever fluxons travel around an AJTJ they leave a clear
signature on the junction current-voltage characteristic (IVC) and therefore
are easily detectable. In fact, as soon as a bias current is fed to the
AJTJ, the fluxons move as magnetic dipoles under the action of the resulting
Lorentz force. The fluxon dynamics in long JTJs is a well known topic and
has received a great deal of both theoretical and experimental attention in
the last few decades. If the external bias is assumed to be uniform over the
junction area, then, as a result of the balance between the externally
supplied power and the internally dissipated power, the fluxons move with a
constant speed: the larger is the external bias the greater is the fluxon
speed, but never exceeding its relativistic limit sets by the Swihart
velocity. The motion direction depends both on the current sign and fluxon
polarity (i.e., whether $n=\pm 1$) for a given bias current. Fluxons having
different polarities travel in opposite directions and are likely to
annihilate when they collide at low speed.

Quantitatively, if a fluxon travels around an AJTJ having a mean
circumference $C$ with a constant speed $v$, then it has a angular speed $%
\omega =2\pi v/C$ and the phase $\phi $ advances of $2\pi $ each period $%
T=2\pi /\omega =C/v$. Therefore, according to the second Josephson equation,
an average voltage $V$ develops across the junction equal to $V=\frac{\Phi
_{0}}{2\pi }\left\langle d\phi /dt\right\rangle =\Phi _{0}/T=\Phi _{0}v/C$
that is proportional to the fluxon velocity. In other words, the presence of
a travelling fluxon sets the junction in a finite voltage state than can be
easily measured on its IVC.

By also changing the bias current through the barrier the voltage drop
changes and a new branch called Zero Field Step (ZFS) appears on the
junction IVC; the ZFS represents the relationship between the applied
Lorentz force (proportional to the bias current) and the fluxon speed
(proportional to the voltage). When $N$ fluxons travel around an AJTJ, the
last expression is easily generalized to give a junction voltage $V=N\Phi
_{0}C/v$. In the last expression $N$ is the total number of travelling
fluxons and can be larger than the winding number $n$ if $F\overline{F}$
pairs are travelling around the annulus. Therefore, we count the number of
travelling fluxons by simply measuring the voltage across the AJTJ.

These properties makes AJTJs very competitive with respect to other solid
state systems proposed to test the Zurek-Kibble mechanism. Our idea is to
perform a large number of N-S transitions on the same AJTJ with no external
current or magnetic field; at the end of each cycle the possible
spontaneously generated fluxons are static. Then we supply an external
current that sets the fluxons (if any) in motion around the annulus and
measure the number of travelling fluxons by a careful inspection of the
junction IVC. Due to the annihilation of a fluxon-antifluxon pair, this idea
works well as long as the chances to spontaneously generate two fluxons are
small.

Fig.1a, Fig.1b and Fig.1c represent the IVC of the same AJTJ with no fluxon
trapped, with one fluxon trapped and with two fluxons trapped, respectively.
We note that with no trapped fluxons the zero voltage current is very large
and only $F\overline{F}$ current steps appear at finite voltage. In the
other two cases the supercurrent is rather small (theoretically it should be
vanishingly small in ideal, pinhole free barriers) and large current
branches can be observed at finite voltages corresponding to the fluxons
and, possibly, $F\overline{F}$ pairs travelling around the junction.

\section{The samples}

High quality $Nb/Al-Al_{ox}/Nb$ JTJs were fabricated on $0.5\,mm$ thick
silicon substrates using the trilayer technique in which the junction is
realized in the window opened in a $SiO$ insulator layer. Details of the
fabrication process can be found in Ref.\cite{VPK}. On each $15\times
24\,\,mm^{2}$ chip four JTJs were integrated, of which three ring shaped
junctions having a mean circumference $C=500\,\mu m$ and a width $\Delta
r=4\,\,\mu m$ and one $4\times 500\,\mu m^{2}$ overlap-type linear junction.
The so called ''idle region'', i.e. the overlapping of the wiring layer onto
the base electrode was about $3\,\mu m$ for all the junctions. The
thicknesses of the base, top and wiring layer were $200$, $80$ and $400\,nm$%
, respectively.

For all samples the high quality has been inferred by a measure of the I-V
characteristic at $T=4.2\,K$ . In fact, the subgap current $I_{sg}$ at $%
2\,mV $ was small compared to the current rise $\Delta I_{g}$ in the
quasiparticle current at the gap voltage $V_{g}$, typically $\Delta
I_{g}>35I_{sg}$, the gap voltage was as large as $V_{g}=2.76\,mV$ and the
maximum critical current $I_{c}$ was larger than $0.55\Delta I_{g}$ for the
overlap type junction. Furthermore, the application of a strong enough
external magnetic field in the barrier plane completely suppressed any
Josephson structures indicating the absence of electrical shorts in the
barrier. It is important to mention that i) no logarithmic singularity has
been observed in the IVCs at low voltages and ii) the temperature dependence
of the critical current was linear as the temperature $T$ approached the
critical temperature $T_{C}$; both these observations assure us that the
junctions are symmetric, i.e. no detectable difference can be assumed
between both the energy gaps $\Delta _{1,2}$ and the critical temperatures $%
T_{C1,2}$ of the junction electrodes $1$ and $2$ in the proximity of the
barrier. The maximum Josephson current density $J_{J} $ was of the order of $%
1\,kA/cm^{2\text{ }}$corresponding to a specific barrier normal resistance $%
\rho _{N}$ of about $200\,\Omega \mu m^{2}$.

Many samples have been measured. For clarity only two will be discussed
here. The geometrical and electrical (at $4.2\,K$) parameters of the two
selected annular junctions on different wafers are listed in Table I. They
have the same geometry (both the base and top electrode have a hole
concentric to the ring,), but differ in the critical current densities,
i.e., in the normalized mean circumferences $C/\lambda _{J}$. The critical
current density has been calculated from the measured quasiparticle current
step, $\Delta I_{g}$, at the gap voltage. The values of the barrier magnetic
thickness $\Lambda =180\,nm$ has been used for numeric calculations. On each
same chip a linear overlap-type junction with the same width, length and
idle region was used in order to measure the junction Swihart velocity $%
c_{0} $ with a geometry in which the effects of the self field are
minimized. The value of $c_{0}=1.4\times 10^{7}m/sec$, due to the effect of
the idle region, is 1.5 times larger than that expected for a bare junction.
This value of $c_{0}$ corresponds to a value of $0.08F/m^{2}$ for the
barrier effective specific capacitance.

\noindent The data in Table I show that both samples are high-quality, long (%
$C\gg \lambda _{J}$) annular JTJs.

\bigskip

\begin{tabular}{lcc}
\hline
\multicolumn{1}{||l}{Sample} & A & B \\ \hline
Mean circumference $C(\mu m)$ & 500 & 500 \\ \hline
Width $\Delta r(\mu m)$ & 4 & 4 \\ \hline
Zero field critical current $I_{o}(mA)$ & 33 & 2.5 \\ \hline
Maximum critical current $I_{\max }(mA)$ & 39 & 2.7 \\ \hline
Gap quasiparticle current step $\Delta I_{g}(mA)$ & 88 & 5.2 \\ \hline
$I_{\max }/\Delta I_{g}$ & 0.45 & 0.52 \\ \hline
Critical current density $J_{c}(A/cm^{2})$ & 3050 & 180 \\ \hline
Josephson length $\lambda _{J}(\mu m)$ & 6.9 & 28 \\ \hline
Normalized mean circumference $C/\lambda _{J}$ & 72 & 18 \\ \hline
Quality factor $V_{m}(mV)$ & 49 & 63 \\ \hline
Normal resistance $R_{N}(m\Omega )$ & 36 & 610 \\ \hline
ZFS1 asymptotic voltage $(\mu V)$ & 51 & 53 \\ \hline
\end{tabular}

\begin{center}
{\bf Table I}
\end{center}

\section{The experimental setup}

\smallskip In order to vary the quenching time in the broadest possible
range, we have realized the experimental setup schematically shown in Fig.2.

\noindent A massive $Cu$ block held to the sample holder by two thin $Cu$
arms was used to increase the system thermal capacity. The chip was mounted
on one side of this block and thermally loosely coupled to it by means of a $%
1\,mm$ thick teflon sheet. On the other side of the $Cu$ block, a
thermoblock consisting of a $50\,\Omega $ carbon resistor and two
thermometers in order to measure and to, if necessary, stabilize the $Cu$
block temperature, was mounted in good thermal contact. Finally a small
sized $100\,\Omega $ resistor, more precisely a surface mount resistor
(SMR), was kept in good thermal contact with the chip by means of a small
amount of vacuum grease.

This system, due to the two heating elements placed in tight and loose
thermal contact with the chip and with the proper choice of the thermal
loads, allowed us to perform the sample quenching over two quite different
time scales. In fact, by means of the resistor in the thermoblock, a long
time scale was achieved by heating the chip through the $Cu$ block and the
teflon sheet; on the contrary, a short current pulse through the surface
mount resistor on the chip, attained much short thermal cycles.

These two completely different quenching techniques provide timescale ranges
that do not overlap, leaving a gap between $0.2$ and $1s$, that would
require a third quenching technique to be filled. We stress that just using
a single sample holder with smaller heat capacity would not give us access
to the same time-scales. We could not get quench times as short as those
from the SMR and, even with a very small exchange gas pressure, could not
get times as long as those with the larger sample holder on using the
mechanical pumps available to us.

The whole system was kept in a vacuum tight can immersed in the $LHe$ bath
at $He$ gas pipeline pressure. The pressure of $He$ gas inside the can could
be varied in order to modify the heat exchange between the chip and the
environment and, in turn, the speed of the sample cooling. A solenoid was
wound around the can to provide a strong vertical magnetic field and
Helmholtz coils were instead placed inside the can to generate a weak
horizontal magnetic field in the barrier plane in order to tune the critical
currents of the annular junctions to their maximum values.

The temperature dependence of the junction gap voltage was exploited to
monitor the temperature of the junction itself during the thermal cycle.
Fig.3a and Fig.3b show the digitally measured $V_{g}(t)$ for sample $A$
current biased on the quasiparticle curve at $17.7\,mA$, (i.e., at about one
fifth of $\Delta I_{g}$ at $T=4.2\,K$), for a slow and fast thermal cycle,
respectively. In the case of Fig.3a, a $100\,mA$ current was fed to the
thermoblock heater for about $4s$ in order to increase the junction
temperature up to its critical value where the sample IVC becomes a straight
line with a slope corresponding to the junction normal resistance $R_{N}$.
In the case of Fig.3b a $20\,V$ high and $4\,ms$ wide, voltage pulse was
applied at the SMR. In the Figs.3 the time origin is arbitrarily set at the
instant we began to feed the heating elements. It is important to observe
that the time scale in Fig.3a is about 50 times larger than that of Fig.3b,
although the curve shapes are quite similar.

In our samples the current jump at the gap voltage is very steep and, at $%
T=4.2\,K$, the voltage changes by less than $1\,\%$ when the current is
changed from $10$ to $90\,\%$ of the total current jump $\Delta I_{g}$ and
by less of $10\,\%$ at $T=8.5\,K$. Furthermore, at $T=4.2\,$and $8.5\,K$ and
for this bias current, the junction voltage was equal to $2.74\,$and $%
1.0\,mV $, respectively. Therefore, assuming that the electrode gap energies
are equal $\Delta _{1}(T)=\Delta _{2}(T)=\Delta (T)$, and that in the $%
4.2-8.5\,K $ range we can neglect the thermal gap smearing, the analytical
expression found by Thouless\cite{thouless} for the gap energy in a
strong-coupling superconductor

\begin{equation}
\frac{\Delta (T)}{\Delta (0)}=\tanh \frac{\Delta (T)}{\Delta (0)}\frac{T_{c}%
}{T},  \label{gap}
\end{equation}

\noindent also applies to the junction gap voltage that is proportional to
it. The dashed lines in Figs.3 indicate the voltage threshold above which Eq.%
\ref{gap} can be used to relate the junction voltage at its temperature,
without any significant loss of accuracy. We like to mention that an
experimental proof of Eq.\ref{gap} in $Nb/Nb$ tunnel junctions was first
evidenced by Broom\cite{broom}.

Fig.4a and Fig.4b show the data reported in Fig.3a and Fig.3b, respectively,
transformed according to Eq.\ref{gap}, assuming for $\Delta (0)$ and $T_{c}$
the values $2.85\,meV$ and $8.95\,K$, respectively, as found by Monaco {\it %
et al.} \cite{cristiano} on similar JTJs. Now the dashed horizontal lines
indicate the temperature threshold below which the temperature time
dependence can be reliably accounted for by our measured data. The large
noise at low temperature is the result of an amplification effect of Eq.\ref
{gap}, according to which the temperature variation corresponding to a given
energy gap variation becomes larger and larger as the temperature becomes
smaller. It would be very complicated to write the proper boundary
conditions for the heat diffusion equation that would correctly model the
full time dependence of the junction temperature. However, for our purposes
we are only interested to the cooling process, and we successfully fit our
data by a simple thermal relaxation equation:

\begin{equation}
T(t)=T_{fin}+(T_{in}-T_{fin})\exp \left( -\frac{t-t_{0}}{\tau }\right)
\label{relaxation}
\end{equation}

\noindent with only two fitting parameters $t_{0}$ and $\tau $, being $%
T_{in} $ and $T_{fin}$ fixed at $8.95$ and $4.15\,K$, respectively. In Eq.%
\ref{relaxation} $t_{0}$ is the time at which $T=T_{in}=T_{c}$ and $\tau $
is the relaxation time which sets the cooling time scale. The fitting curves
are shown by the thick dashed lines in Fig.4a and Fig.4b, and correspond to
a thermal relaxation time $\tau $ equal to $3.6$ and $0.073\,s$,
respectively. The quenching time $\tau _{Q}$ can be obtained from its
definition:

\begin{equation}
\frac{T_{C}}{\tau _{Q}}=-\frac{dT}{dt}\mid _{T=T_{C}}  \label{tau_q}
\end{equation}

\noindent giving $\tau _{Q}=\tau T_{C}/(T_{in}-T_{fin})$. For a thermal
relaxation from the junction critical temperature down to the helium bath
temperature we get $\tau _{Q}\simeq 1.9\,\tau $. Eventually, Fig.5 display
the values of the quenching times obtained with the process described above,
both for the fast (black squares with right vertical scale) and slow (red
circles with left vertical scale) cooling processes and for different values
of the $He$ pressure inside the can. We observe that, by changing the
exchange gas pressure and using the two techniques, the quenching time can
be changed over a quite large range starting from tenths to tens of seconds.
At the end of this paper we will discuss how it is possible to extend this
range in both directions; however, as we will show in next section, this
range has shown to be large enough for our purposes.

\section{The measurements}

Quenching experiments were carried out in a double $\mu $-metal shielded
cryostat and the transitions from the normal to the superconducting states
were performed with no current flowing in the heaters and the thermometers.
Both the junction voltage and current leads were shorted during all the
thermal cycle. Furthermore, the heat supplied to the sample was such that
the maximum temperature reached by the junction was made slightly larger
than its critical temperature, say at about $10\,K$, in order to make sure
that also the bulk electrode critical temperature ($T_{C}\simeq 9.2K$) was
overcome. In this case, according to Eqs.\ref{relaxation} and \ref{tau_q},
the value of the quenching time results in a correspondingly smaller $\tau
_{Q}\simeq 1.7\,\tau $. Due to the approximation made by using Eq.\ref
{relaxation} and to the experimental uncertainty in the knowledge of the
maximum temperature during each thermal cycle, the value of the quenching
times has been determined with an overall accuracy as large as $5\%$. For
each value of the quenching time, in order to estimate the trapping
probability, we have carried out a set of 300 thermal cycles and at the end
of each cycle the junction IVC was inspected in order to ascertain the
possible spontaneous trapping of one or more fluxons.

As we shall see later, the AJTJs are such that the ZK causal length ${\bar\xi%
}> C$ by an order of magnitude when $\tau_Q = 1\thinspace s$. Increasing and
decreasing $\tau _{Q}$ by an order of magnitude changes ${\bar{\xi}}$ by
less than a factor of two. Thus the probability of finding a single fluxon
after a quench is small. In the following we will focus our attention only
on the probability $P_{1}$ to trap just one fluxon, although a few times we
found clear evidence of two and, more seldom, three homopolar fluxons
spontaneously trapped during the N-S transition. However, these events were
too rare to be statistically significant.

Experimentally, we define $P_{1}$ as the ratio between the number of times
in which at the end of the thermal cycle the junction IVC looks like that
shown in Fig.1b, i.e. with a tiny critical current and a large first ZFS,
and the number of attempts. It is worth to mention here that in the case of
simultaneous trapping of a fluxon and an antifluxon, they would annihilate
and leave no track of their formation. Therefore, our definition of $P_{1}$
is not rigorous, but it is reasonable as far as the chances to trap two
(homo or heteropolar) fluxons are negligibly small. Similarly, an IVC
similar to that shown in Fig.1b could be the result of the simultaneous
trapping of two fluxons and one antifluxon, or the other way around, but
this event is less likely in our experimental situation. For the sake of
completeness, it must be added that, in some cases, the IVC displayed either
a depressed critical current without ZFS structures or a ZFS with an
enhanced critical current. We explain them as due to the trapping of
Abrikosov vortices in the junction electrodes and nearby the barrier and we
did not take in any account the occurrence of such events, since it is not
known if and how the vicinity of Abrikosov vortices influences the fluxon
formation.

\section{The results}

When ${\bar{\xi}}>C$ we estimate the probability of finding a fluxon in a
single quench to be 
\begin{equation}
P_{1}\simeq \frac{C}{\bar{\xi}}=\frac{C}{\xi _{0}}\bigg(\frac{\tau _{Q}}{%
\tau _{0}}\bigg)^{-\sigma },  \label{P1}
\end{equation}
where, from Eq.(\ref{JTJ}), $\sigma =0.25$.

Fig.6 shows on a log-log plot the measured probability $P_{1}$ of a single
fluxon trapping obtained by quenching the sample $A$ 300 times for each
value of the quenching time $\tau _{Q}$ changed by varying the exchange gas
pressure and by using both the fast and slow quenching techniques. We
observe that the points are quite scattered meaning that the data are
statistically poor. Further, for the reasons given earlier there is a gap
between fast and slow quenches. Nonetheless, we have clear evidence that i)
the trapping of a fluxon occurs on a purely statistical basis being
identical the conditions for each thermal cycle and ii) the probability to
trap one fluxon is larger when the transition is performed at a faster speed
(smaller quenching time) in a qualitative accordance with the causality
principle. More precisely, we can distinguish the point to the lower right
of the graph corresponding to the slow cycle process which gives in the
average a probability to spontaneously trap a fluxon once every fourteen
attempts, and the cloud of data to the upper left corresponding to the
impulsive junction heating and giving an average probability of one
successful event every about six attempts. This suggests that possible
temperature gradients induced by the SMR are not an important source of
systematic error, since such gradients, with their slowly moving profiles,
have a tendency to {\it reduce} defect production\cite{volovik}.

Regardless of the data spread, as suggested if Eq.\ref{P1} holds true, we
attempted to fit the data with an allometric function $P_{1}=a\cdot \tau
_{Q}^{-b}$ with $a$ and $b$ being free fitting parameters. We found that the
best fitting curve, shown by the solid line in Fig.6, has a slope $b=0.27\pm
0.05$. Such a value of $b$, although affected by a $20\%$ uncertainty, is in
good agreement with the forth root square dependence expected for symmetric
junction.

For the coefficient $a$ we found the best fitting value of $0.1\pm 10\%$ ($%
\tau _{Q}$ in seconds). This is to be compared with the predicted value of $%
C\tau _{0}^{1/4}/\xi _{0}\,$. Sample $A$ had a circumference $C=500\,\mu m$.
Its effective superconductor thickness was $d_{s}\approx 250$ $nm$. At the
final temperature $T_{fin}=4.2\,K$, the critical current density was $%
J_{c}(T_{fin})=3050\,A/cm^{2}$ and the Josephson length was $\lambda
_{J}(T_{fin})=6.9\,\mu m$. From this, and $c_{0}$ given earlier, we infer
that $\xi _{0}\approx 3.8\,\mu m$ and $\tau _{0}\approx 0.17\,ps$. This then
gives $C\tau _{0}^{1/4}/\xi _{0}\approx 0.08\,s^{1/4}$, in good agreement
with the experimental value of $b$, given the fact that we only expect
agreement in overall normalization to somewhat better than an order of
magnitude level. At this level, such a result is immune to systematic error
in one or other of the measurement processes. After the problems (discussed
below) of the experiments discussed in \cite{lancaster,lancaster2,technion}
to find (reliable) defects at expected densities, if at all, our experiment
shows that the ZK estimate remains sensible.

Similar measurements have been carried out for sample $B$. Although not in
contradiction with (\ref{xibar}), the results were affected by a data
scattering even larger than that found for sample $A$ (shown in Fig.6). This
is due to a much smaller normalized length which, according to (\ref{P1}),
translates in a expected probability $P_{1\text{ }}$, for a given $\tau _{Q}$%
, about four times smaller (since $C\tau _{0}^{1/4}/\xi _{0}\approx
0.02\,s^{1/4}$ for this sample), far too small too get statistically
significant data in reasonable times considering that these measurements are
both very time and $LHe$ consuming. In order to have data comparable with
those of sample $A$, sample $B$ would have required a quenching times $4^{4}$
times larger. However the roughly measured probability $P_{1\text{ }}$of 1
fluxon every 50-100 attempts is in fairly good agreement with the expected
value. This shows that our fluxons are not spurious byproducts of the
measurement mechanism.

\section{Comments, Future Experiments and Conclusions}

We consider this experiment to give a strong confirmation of the
Zurek-Kibble predictions. We said in the introductory section of this paper
that condensed matter experiments to test the ZK predictions had given mixed
results, and it is interesting to put this experiment in that context.

Prior to our experiment, seven other experiments had been performed to test (%
\ref{xibar}), five with fixed $\tau _{Q}$\cite
{helsinki,grenoble,lancaster,lancaster2,technion}, two with variable $\tau
_{Q}$\cite{pamplona,florence}. [ In addition, the experiment cited earlier%
\cite{Roberto} on JTJs by two of us (R.M and J.M) was compatible with (\ref
{JTJ}), although it had not been performed with a test of (\ref{JTJ}) in
mind. It was this that motivated the experiment described here.]

Of those experiments with fixed $\tau _{Q}$, two were experiments\cite
{helsinki,grenoble} on superfluid $^{3}He-B$, which rely on the fact that,
when it is bombarded with slow neutrons, energy is released which leads to a
hot spot, with temperature $T>T_{c}$, in the superfluid which then cools
below $T_{c}$. This leaves behind a tangle of vortices, the topological
defects in this system, whose density can be measured. Since $\tau _{Q}$ is
fixed by the nuclear process, it is not possible to confirm the predicted
value $\sigma =1/4$. However, with only a single data point conflating both
normalisation and $\sigma $ both experiments are highly compatible with (\ref
{xibar}).

The remaining experiments with fixed $\tau _{Q}$ were two\cite
{lancaster,lancaster2} on superfluid $^{4}He$, and one\cite{technion} on
high temperature superconductors (HTSC).

In principle, the $^{4}He$ experiments\cite{lancaster,lancaster2}, which use
a pressure quench with a varying timescale $\tau _{Q}$ to implement the
transition, could have allowed for a more complete test, in this case to
confirm $\sigma =1/3$ (after renormalisation group rescaling). Yet again,
vortices are the relevant defects. In practice, the most reliable published
experiment\cite{lancaster2} sees no vortices. In this context, the vortices
seen in an earlier $^{4}He$ experiment\cite{lancaster}, at levels compatible
with (\ref{xibar}), were most likely an artefact of the experimental setup.
Further experiments on $^{4}He$ are underway.

The fifth experiment\cite{technion}, on HTSC, measures total flux through a
surface i.e. the variance in the topological charge, carried in this case by
the Abrikosov vortices. The vortex separation of (\ref{xibar}) can be
converted into a prediction for the variance, but no flux is seen in
contradiction with this prediction, despite the phase separation that is a
prerequisite for the result being seen elsewhere\cite{carmi}. There is no
obvious explanation of this null result. An attempt to take gauge fields
into account fully\cite{rajantie} shows that there is an additional
mechanism for vortex production in the thermal fluctuations of the magnetic
field but, as yet, this seems insufficient to explain the result. Such a
mechanism will not apply to the JTJs considered in our experiment.

These early experiments have either provided one data point for (\ref{xibar}
), or have been null for whatever reason. Two subsequent experiments have
permitted varying quench rates and so an estimate for $\sigma $. The most
recent\cite{pamplona} involves the B$\acute{e}$nard-Marangoni
conduction-convection transition, in which a homogeneous conduction state is
broken into an hexagonal array of convection lines on heating. The defects
here are not associated with the line zeroes of an order parameter field,
and the viscosity-dependent $\sigma $ does not match the ZK prediction, most
likely for that reason. The more relevant experiment\cite{florence} is
carried out in a non-linear optical system, with complex beam-phase the
order parameter, satisfying a time-dependent Ginzburg-Landau equation with
drift. There has been much numerical analysis\cite{zurek3} of time-dependent
Ginzburg-Landau systems, which show agreement with the ZK predictions for
scaling exponents. The control parameter in \cite{florence} is not the
temperature, but the light intensity. Increasing it leads to pattern
formation (defects) at a critical value. The predicted scaling parameter $%
\sigma =1/4$ is recovered to good accuracy as $\sigma _{exp}=0.25\pm 0.02$,
but agreement with normalisation is not stated.

Given this relatively poor success rate in confirming (\ref{xibar}) we are
considering a further experiment to measure the ZK scaling exponent, this
time with manifestly non-symmetric AJTJs. In \cite{KMR} and \cite{MRK} we
observed that it should have been difficult to make JTJs truly symmetric, as
those used here. However, in \cite{KMR} and \cite{MRK} we had not
appreciated how the difference between symmetric and marginally
non-symmetric JTJs is smeared by the proximity effect of the $Al$ within the
insulating layer. Significantly non-symmetric JTJs require different
fabrication techniques, but the value of $\sigma $ inferred from the same
causal arguments is $\sigma =1/7$, very different from the value of $1/4$
that we tested above. The data from our experiment is incompatible with $%
\sigma =1/7$. This does suggest that a further experiment, with markedly
non-symmetric JTJs, should be performed.

Our experiments have demonstrated that quenching time of the order of 1
second give a rather large probability to trap one fluxons on AJTJs having a
very large normalized length. However, very long junctions mean very large
critical current densities that, in turn, require Josephson barriers so thin
that their quality and uniformity is often spoiled; furthermore, in most
cases, applications require intermediate length junction or even small
junction. For these reasons, it would be highly desirable to compensate the
reduced junction length with an increased quenching rate, as it is suggested
by the findings for sample $B$. Therefore, we like to conclude the paper
with some comments on the possible technical improvements that would allow
to test the ZK predictions over a broader quenching time range. Firstly, $%
\tau _{Q}$ can be trivially heightened by increasing the $Cu$ block thermal
capacitance. On the contrary in order to lower the quenching time, that is
to make the $N-S$ transition faster, it is needed to resort to new
techniques since the maximum power that can be dissipated by the surface
mount resistors sets an obvious lower threshold on $\tau _{Q}$. \smallskip
One possible way to reach this goal is to perform the junction thermal cycle
by means of light pulses. Light dissipates inside the superconducting
electrodes, but not in the substrate providing a local junction heating that
will relax much faster to the background temperature. We estimate that, by
using a properly focussed pulsed light beam, the quenching time scale can be
reduced to the $\mu s$ range.

\smallskip

\section*{ACKNOWLEDGMENTS}

The authors thank L. Filippenko for the sample fabrication and V.P.
Koshelets for useful discussions. R.R. thanks the University of Salerno for
hospitality. This work is also supported by the COSLAB programme of the
European Science Foundation.

\section{\bf Table and Figure Captions}


\begin{itemize}
\item[Table I]  {Geometrical and electrical parameters of two selected
annular Josephson tunnel junctions at 4.2 K.}

\item[Fig. 1]  Low voltage part of the experimentally measured
current-voltage characteristics of the same annular Josephson tunnel
junction a) without trapped fluxons, b) with one trapped fluxon, and c) with
two trapped fluxons. For each current branch the corresponding number of
travelling fluxons $F$ and fluxon-antifluxon pairs $F\overline{F}$ is
indicated.

\item[Fig.2]  Sketch (dimensions are not to scale) of the cryogenic insert
developed to perform the junction thermal cycles with a time scale changing
over a broad range.

\item[Fig.3]  Digitally measured time dependence of the junction gap voltage
during the a) ''slow'' and b) ''fast'' thermal cycle. For these measurements
the junction was biased at about 1/5 of the gap quasiparticle current step.
The horizontal dashed lines indicate the voltage threshold above which Eq.%
\ref{gap} can be used to relate the junction voltage at its temperature.

\item[Fig.4]  Time dependence of the junction temperature during the a)
''slow'' and b) ''fast'' thermal cycle obtained from the data of Figs.3
transformed according to Eq.\ref{gap}. The horizontal dashed lines indicate
the temperature threshold below which the temperature time dependence can be
reliably accounted for by our measured data. Furthermore, the thick dashed
lines are best fitting curve of the cooling process according to the thermal
relaxation expression Eq.\ref{relaxation}.

\item[Fig.5]  Quenching time $\tau _{Q}$ as a function of the $He$ gas
pressure inside the vacuum tight can. By changing the pressure of the
exchange gas inside the can, the system thermal constants are varied. The
solid squares refer to the left vertical scale while the solid circles
refers to the right vertical scale.

\item[Fig.6]  Log-log plot of the measured probability $P_{1}$ to trap one
fluxon versus the quenching time $\tau _{Q}$. The solid line is the best
fitting curve found assuming a power law dependence as suggested by Eq.\ref
{xibar}. To a good degree of approximation the fit is in agreement with a
forth root square dependence as expected for symmetric annular Josephson
tunnel junctions.
\end{itemize}

\end{document}